\documentclass[%
 reprint,
 amsmath,amssymb,
 aps,
pra,
]{revtex4-2}

\usepackage{graphicx}
\usepackage{dcolumn}
\usepackage{bm}

\begin{document}

\preprint{APS/123-QED}

\title{Mechanical Long Baseline Differential Gradiometers as Low Frequency Gravitational Wave Detectors}

\author{Enrico Calloni}
\email{enrico.calloni@na.infn.it}
\author{Annalisa Allocca}
\author{Antonino Chiummo}
\author{Rosario De Rosa}
\author{Luciano  Errico}
 \author{Marina Esposito}
 \author{Edoardo Imparato}
 \author{Bruno Mantice}
 \author{Luigi Rosa}
 \author{Alessandra Ruggiero}
 \author{Valeria Sequino}
\author{Daniela Stornaiuolo}
\author{Vittorio Tortorella}
  \author{Lucia Trozzo}
  
 \affiliation{
 Università di Napoli Federico II  
 Via Cinthia, I-80124 Naples, Italy \\
}%
\affiliation{
INFN, sez. Napoli,  
 Via Cinthia, I-80124 Naples, Italy\\
}%

 \author{Paolo Ruggi}
\affiliation{
European Gravitational Observatory (EGO), I-56021 Cascina, Pisa, Italy\\
}%

\date{\today}

\begin{abstract}

We present a new differential mechanical gradiometer for the detection of low-frequency Gravitational Waves. The frequency range is 0.05 – 1 Hz, a frequency gap not covered either by future space-based detectors such as LISA or by ground-based observatories such as Einstein Telescope or Cosmic Explorer.
 The proposed detection principle is similar to antennas based on torsion pendulums but solves the problem of physical confinement of these antennas by operating vertically and by having a counterweight at one end of each bar and a mass suspended from a long wire at the other. With this configuration, we enlarge the gravitational force acting on the system  \textit{without} changing the moment of inertia of the system, so that we move from a signal $\Delta \theta$ of the order of $\Delta \theta = h$, where h is the amplitude of the gravitational wave, to a signal of the order $\Delta \theta = h\frac{L}{D}$, where D is the length of the arm and L is the length of the wire suspending the test mass.  This configuration is 
 a further evolution of the recent development of tiltmeters and balances with double suspended arms and interferometric read-out, where the main working principles are already tested. The expected sensitivity will be discussed with respect to the proposed parameters and the present technology. 
\end{abstract}

\maketitle


\section{\label{sec:introduction}Introduction}

By a mechanical gravitational wave detector, we mean a continuous mechanical system in which at least one observable is modified by the interaction with a gravitational wave. This includes the first historic resonant Weber bar antennas, which detect the oscillation amplitude of the fundamental longitudinal mode, their evolution into cryogenic bar antennas, the proposals for more complex geometric shapes, such as crosses or spheres, and the use of torsion pendulums \cite{Weber:1960zz,Astone:1997gi,Aguiar:2006va,Takano:2024vht}.

A crucial disadvantage of all the mechanical detectors has been, and still is, their physical size, on the order of a few meters. Since the gravitational wave effect is proportional to the physical size of the detector, mechanical detectors have gradually given way to interferometers, whose physical size is on the order of a few kilometers.
As is well known, it is these detectors that paved the way for the discovery of gravitational waves and their systematic study \cite{LIGOScientific:2016aoc,LIGOScientific:2017vwq,LIGOScientific:2021usb,LIGOScientific:2025hdt}.

An interesting point regarding the future detection of gravitational waves is the frequency gap left open between ground-based interferometers, both current and future, and space-based interferometers. As is known, current ground-based interferometers have a lower detection limit around 10 Hz, which will extend to a limit of 2-3 Hz in the future observatories \cite{Punturo:2010zz,Sathyaprakash:2012jk,Hall:2020dps,Hall:2022dik}. Space-based interferometers, like LISA \cite{Robson:2018ifk}, conversely, will have best sensitivities around tens of mHz, so it is expected that the frequency range from one hundred mHz to a few Hz will be covered by new detectors or will risk remaining unexplored.
In this frequency region, various types of detectors are being designed or are undergoing initial experimental exploration. These include experiments based on atomic interferometry, such as MIGA \cite{Canuel:2017rrp}, MAGIS-100 \cite{Coleman:2018ozp}, and AION 
\cite{Aion}, 
aimed at future large-scale detectors, such as ELGAR \cite{Canuel:2019abg}, ZAIGA \cite{Zhan:2019quq}, AEDGE \cite{AEDGE:2019nxb}, and MAGIS \cite{Coleman:2018ozp}, and mechanical detectors, particularly torsion pendulums.
In this frequency range, torsion pendulums are certainly of interest \cite{Takano:2024vht}. These instruments are ideal for detecting small forces on macroscopic objects, with typical observational bandwidth in the frequency range just above a few mHz, and with projected sensitivities comparable with atomic interferometry detectors.
For this purpose, however, the typical size of the order of a few meters imposes stringent limits on the detector's characteristics and difficulties in achieving the sensitivity required for the observation of gravitational waves. 
In this context, we propose as a new approach, the use of mechanical balances suspending a test mass with a suitable long wire, or equivalently with an elastic joint and a rigid negligible-mass rod, so to extend the physical size of the detector {\it without} enlarging the inertial moment of the detector, so as to amplify the effect of the gravitational wave on the arm's tilt signal.
Experimentally, this approach is made possible by the recent development of instruments such as tiltometers and mechanical balances. In these instruments, the arm is suspended to the ground with thin joints, and for balances, it is capable of suspending additional test masses by wires \cite{Venkateswara:2014nra,Allocca:2021kua,Allocca:2024lao,Allocca:2025ebf}. 
The rotation is in the vertical plane. Their scheme makes it possible to extend the physical size of the detector naturally, simply by extending the length of the suspension wire for the test masses, as shown in Fig.~\ref{fig:differentialGradiometer}.
Technically, our proposal is a large-base mechanical gradiometer, and in the configuration proposed here, a pair of such gradiometers, differentially coupled, forms the complete detector. As we will show, compared to a torsion pendulum, where the differential rotation signal of the arms, for optimal polarization, is $\Delta \Theta = h$, where h is the amplitude of the gravitational wave, the differential gradiometer signal becomes $\Delta \Theta = h \frac{D}{L}$ where D is the length of the wire suspending the test mass and L is the length of the gradiometer arm. Since in realistic configurations, D can be at least on the order of hundreds of meters and L on the order of a meter, this results in the possibility of increasing the sensitivity to gravitational signals.
In this article, after a brief review of the operating principle, we will illustrate a realistic configuration, clarifying the achievable sensitivities with respect to present technologies.

\section{\label{sec:Operating Principle}Operating Principle}

The operating principle is based on the equation, written in the proper detector frame, which links the gravitational force $F_{i} $ acting on a mass m, placed at position $\xi^{k} $, and the gravitational wave described in the Transverse Traceless gauge, $h_{ij}^{TT}$ \cite{Maggiore}
\begin{equation}
F_i = \frac{m}{2}\ddot h_{ij}^{TT} \xi^{j}
\end{equation}

The equation written, as usual, in Fourier transform, becomes \begin{equation}
\tilde{F}_i = -\omega^2\frac{m}{2}\tilde{h}_{ij}^{TT} \xi^{j}
\end{equation} 
where $f = \frac{\omega}{2\pi}$ is the frequency of the gravitational wave.
For simplicity, we focus on a gravitational wave with a propagation direction equal to the direction of the gradiometer's rotation axis (namely the z axis). We choose the origin of the proper detector frame at the gradiometer's rotation point, with axes parallel and perpendicular to the gradiometer arm. Finally, we assume the wave is "+" polarized, monochromatic of frequency of $f = \frac{\omega}{2\pi}$ and amplitude $h$.
Under these assumptions, the previous equation can also be written in Fourier transform, 
\begin{equation}
\tilde{F}_i = -\omega^2\frac{m}{2}h{\bf e}_{ij}^{+} \xi^{j}
\end{equation} 
where h is the amplitude of the gravitational wave, $f = \frac{\omega}{2\pi}$ is the frequency of the gravitational wave, and ${\bf e}$ is the "+" polarization matrix.

\begin{figure}
    \centering
    \includegraphics[width=0.5\linewidth]{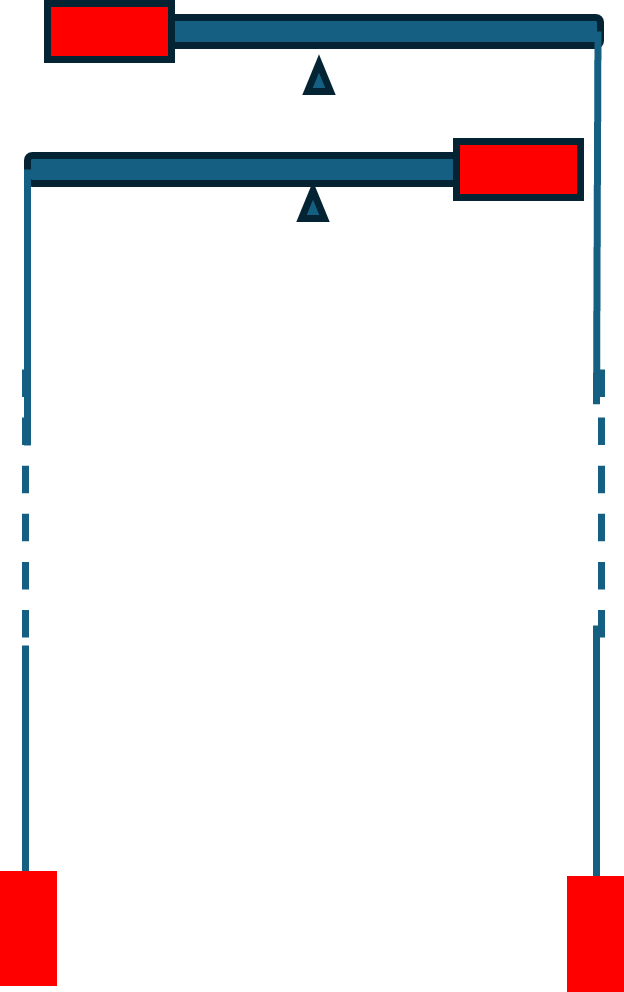}
    \caption{Detector schematic. Each arm is a gradiometer whose signal is the difference in vertical gravitational field between the position of the suspended mass and the counterweight. The two arms, due to the arrangement of the masses, generate a differential signal of angle variation. This signal is read interferometrically.}
    \label{fig:differentialGradiometer}
\end{figure}
To calculate the effect of a gravitational wave, we focus on the gravimeter I, having masses $m_1^{(I)} = m_2^{(I)} = m$, as shown in the figure \ref{fig:differentialGradiometer}. (The effect of the gravitational wave on the second gravimeter generates an opposite rotation of the arm, doubling the signal.)

The torque exerted by the gravitational wave acting on the mass $m_1^{(I)}$ on the left is zero: indeed, it is given by

\begin{equation}
\tau_z = \frac{m}{2} e_{3ki}\xi^{j}\xi^{j}\ddot h_{ij}^{+} 
\label{torque1}
\end{equation}

and it is equal to zero because $\xi^{2} = 0 $ and $h_{ij}^{+}$ is diagonal (please notice that here "2" is the index).

The force exerted by the gravitational wave on the suspended mass $m_2^{(I)}$ has both a vertical and horizontal component. This second component is much smaller than the first, because $\xi^{1} = L/2$ while $\xi^{2} = D$. Furthermore, when subjected to a horizontal force, the wire bends but, being attached to the arm (very near) at the height of the center of mass, it does not exert a torque on the arm, making this component, in fact, negligible in the calculation of the effect.

To calculate the effect of the vertical component, it is advisable to explicitly write the equations of motion related to the coupling of arm tilt "$\theta$", the vertical position of the suspended mass "y", and the force "$F_z = \frac{m}{2}\ddot{h}y$" exerted by the gravitational wave, always evaluated in the approximation of $\theta << 1$, with the aim also  to clarify further the assumptions made: 

\begin{equation}
\left \{ \begin{array}{rl}
 I \ddot{\theta} = mg\frac{L}{2}-k_{\tau}\theta - k_w \left(\frac{L}{2}\theta - y - D\right)\frac{L}{2} \\

m\ddot{y} = -mg + k_w \left(\frac{L}{2}\theta - y - D\right) + \frac{m}{2}\ddot{h}y
\end{array}
\label{torque}
\right.
\end{equation}

where I is the moment of inertia of the arm+mass$_1$, D is the length of the wire, L is the length of the arm, $k_{\tau}$ is the elastic constant of restoring torque and $k_w$ is the vertical spring constant of the wire.

We just note that these equations are equivalent to writing the Lagrangian 

\begin{equation}
\begin{split}
\mathcal{L} = \frac{1}{2}I\dot{\theta}^2  +   \frac{1}{2} m\dot{y}^2  +mg\frac{L}{2}\theta - mgy + \\
- \frac{1}{2} k_{\tau}\theta^2 - \frac{1}{2}k_w\left(sin\theta\frac{L}{2} - y - D\right)^2 + m\frac{\ddot h}{4} y^2 
\end{split}
\end{equation}

where the gravitational wave acts as an external field along the "y" axis and where we point out that the kinetic term of the suspended mass is related to the tilt $\theta$ by the lever length L/2. Returning to the equations of motion, multiplying the second by L/2 and adding we get:

\begin{equation}
I \ddot{\theta} + m\ddot{y} \frac{L}{2} = - k_{\tau}\theta - \frac{mL}{4}\ddot{h}y    
\end{equation}

In the frequency band of interest for the gravitational wave detector, i.e. for $2 \pi f << \sqrt{K_w/m}$, with the wire bending and remaining vertical, the y position of the mass follows the tilt of the arm and $y = L/2sin\theta -D$ and the previous equation becomes

\begin{equation}
\left( I  + m\frac{L^2}{4}\right)\ddot{\theta} = -k_{\tau}\theta + \frac{mL}{4}\ddot{h}y 
\end{equation}

So, in this regime, the equation of motion is equivalent to a rotating system having the Inertial moment $I_E = I + m\frac{L}{4}^2$ equivalent to a system having both the mass located at the distance $L/2$ from the center of rotation, but forced, in the suspended one, by the gravitational wave at the much longer physical distance D:
\begin{equation}
I_E\ddot{\theta} =  -k_{\tau}\theta + \frac{mL}{4}\ddot{h}D
\label{finalequation1}
\end{equation}

As a further clarification, we note that in the case in which the second mass was not suspended with a wire but anchored to a rod of length D, the Lagrangian would become

\begin{equation}
\begin{split}
\mathcal{L} = \frac{1}{2}\left[I + m\left(\frac{L^2}{4} + D^2\right)\right]\dot{\theta}^2  -  \frac{1}{2} k_{\tau}\theta^2 + m\frac{\ddot h}{4} y^2 
\end{split}
\end{equation}

we use this expression to underline how the kinetic term $ \frac{m}{2}\left( \frac{L^2}{4} + D^2\right)\dot{\theta}^2$ would imply a response of the system with an high moment of inertia,  which would greatly reduce tilt effect. 
\footnote{Let us notice that for a rigid system this would not be the optimal polarization: it can be shown that for optimal polarization, the signal would be proportional to $D^2$, so that one could return to the usual relation $\ddot\theta = \frac{1}{2}\ddot h$, valid for the single arm of a torsion pendumlum}.  

Rewriting the equation \ref{finalequation1} in the physical condition in which the moment of inertia of the rod is much smaller than the moment of inertia of the masses m, so that $I_E \approx 2m\frac{L^2}{4}$we obtain:

\begin{equation}
I_E\ddot{\theta} =  -k_{\tau}\theta + \frac{I_E}{2}\ddot{h}\frac{D}{L}
\label{finalequationApprox}
\end{equation}

The result of the equation \ref{finalequationApprox} can be compared with the analogous equation for a single arm of a torsion pendulum. As well remembered, for example, in \cite{Takano:2024vht}, in the optimal case of negligible transverse moments of inertia compared to the moment $I_z$ and optimal polarization, the equation for a single rigid arm becomes:

\begin{equation}
I_z\ddot{\theta} =  -k_{\tau}\theta + \frac{I_z}{2}\ddot{h}
\label{finalequation}
\end{equation}

By comparing these two equations, it can be appreciated that suspending the mass from a wire or an elastic joint increases the gravitational signal by the factor D/L.

\section{\label{sec:A realistic case}A case study with realistic parameters}

This study, as mentioned, is motivated by the growth in balance technology and in the several technologies related to gravitational wave interferometers. This validates the description and simulation of wires up to tens of meters long, with suspended masses of the order of hundreds of kg and mechanical elements of the order of a meter. Our non-extreme extrapolation is therefore based on a system with 300 kg masses, with D equal to about 150 m, and a 2 m long arm with a mass of 30 kg. Notice that 300 kg is approximately the mass of the future Einstein Telescope payload \cite{Punturo:2010zz}, 150 is the depth of the a present cavern on a seismically quite Sardinian site \cite{Naticchioni} (Sos Enattos, Sardinia Island, Italy), and the arm length is slighlty higher than that of the Archimedes experiment, which has actually demonstrated the feasibility of a double suspended arm balance with interferometric read-out \cite{Allocca:2024lao,Allocca:2025ebf}. 
Experimentally, to simplify the actual realization, the initial part of the wire, where the bending occurs, can be replaced with an elastic joint. The joint can eventually be connected to a more rigid element, such as a long rod or a non-thin metal wire, which is attached to the final mass. In this configuration, considering that the elastic restoring force adds to the restoring force of the joints suspending the arm, the joint can be more easily dimensioned to demonstrate the achievement of the desired overall resonant frequency, in particular, maintaining low losses.  

In fact, the joints envisaged in this preliminary analysis are 150 mm wide, 0.1 mm thick, 30 mm high, and made of sapphire. The resulting resonant frequency is 6.7 mHz, the expected losses are $\phi = 3.6*10^{-9}$. 

Regarding these particular joints,  the estimation of the mechanical behavior was carried out both analytically 

\footnote{ Let's recall the main laws for a plate with a load of mass m: let E be the Young's modulus, s the thickness of the joint, b the width, and M the mass of the suspended load. We have that $\lambda = \sqrt{\frac{JE}{Mg}}$, $J = \frac{bs^3}{12}$ and the restoring torque constant 
$\tau = Mg\lambda(1 + i\phi)$. In our case, the three joints are equal, the first two suspending the whole mass $M_T = 2m + m_b$ and the third one suspending the mass m. The joint area is chosen to remain at half the breaking load. The resonant frequency is thus 
$ f_0 = \frac{1}{2\pi} \sqrt{\frac{2\tau_a + \tau_w}{I}}$, where I is the total inertial moment and $\tau_a,w$ are the restoring torque constant of the joints suspending the arm and the mass, respectively. }
and through a finite element program, called Octopus, developed within the Virgo collaboration \cite{Ruggi:2025vuz}
Further analysis, to optimize the shape and size of the joint or wires, may lead to further improvement of these results.

From an experimental perspective, a second important point is also to generate an immediately differential signal: as already described, this is the goal of the second gradiometer, symmetrical to the first, placed a few tens of cm above the first. This configuration is proposed primarily because the passage of a gravitational wave generates a differential signal about $\delta\theta = h\frac{L}{D}$, but also because it tends to minimize common environmental signals.
Furthermore, this arrangement of the arms makes it easy to implement an interferometric read-out of the differential angle, according to a scheme already tested, for example, in \cite{Allocca:2024lao}. 

A list of the main parameters is shown in table \ref{tab:parameters}. The suspension resonance frequency $f_0 = 6,7$ mHz is reputed sufficient for a first analysis, considering that frequencies of around 10 mHz have been obtained in tiltometers with similar configurations, and it seems realistic to reach this frequency value. The frequency of the first internal mode $f_I = 1100 Hz$ has been calculated using a commercial finite element simulation program (Inventor), considering the arm made of aluminum and the test masses in tungsten. The quality factors, in the range of a few thousand, are compatible with carefully machined metallic systems.

\begin{table}
    \centering
    \begin{tabular}{ccc}
        Physical Parameter & Value & Unity\\
         Test mass & 300 & kg\\
         Wire length & 150 & m\\
         Arm mass & 30 & kg\\
         Suspension resonant frequency& 6.7 & mHz\\
         Internal Resonant frequency& 1100 & Hz \\
         Suspension loss angle& 3*10{-9} & adim \\
         Internal modes loss angle & 3*10{-4} & adim \\
         Laser input power & 20 & W\\
         Rock density & 2700 & kg$/m^3$\\
         Speed of long. waves & 6000 & m/s\\
    \end{tabular}
    \caption{Values of the manin parameter of the detector}
    \label{tab:parameters}
\end{table}

\begin{figure}
    \centering
    \includegraphics[width=1.0\linewidth]{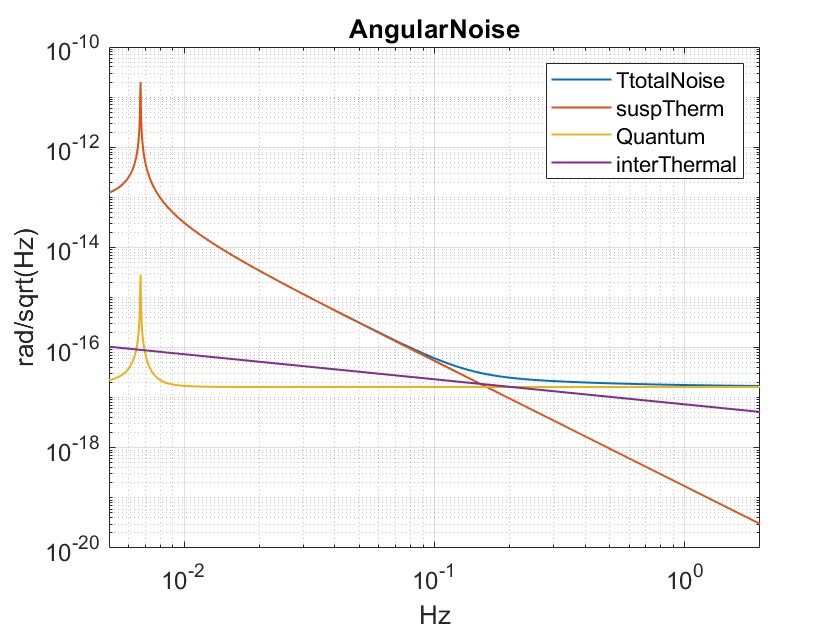}
    \caption{Angular sensitivity as limited by the fundamental noises}
    \label{fig: AngularNoise}
\end{figure}

With these choices, the fundamental noises, shown in the Fig.~\ref{fig: AngularNoise}, were considered.

This angular sensitivity, although not yet achieved by instruments at least in principle similar to the one proposed, such as torsion pendulums or the Archimedes balance (whose sensitivity around 1 Hz is a few 
$ 10^{-12}$ $ rad/\sqrt{Hz}$), does not appear prohibitive and, with the continuous progress of these instruments, and considering the proposed parameters, it is possible not to consider it unattainable.

\begin{figure}
    \centering
    \includegraphics[width=1.0\linewidth]{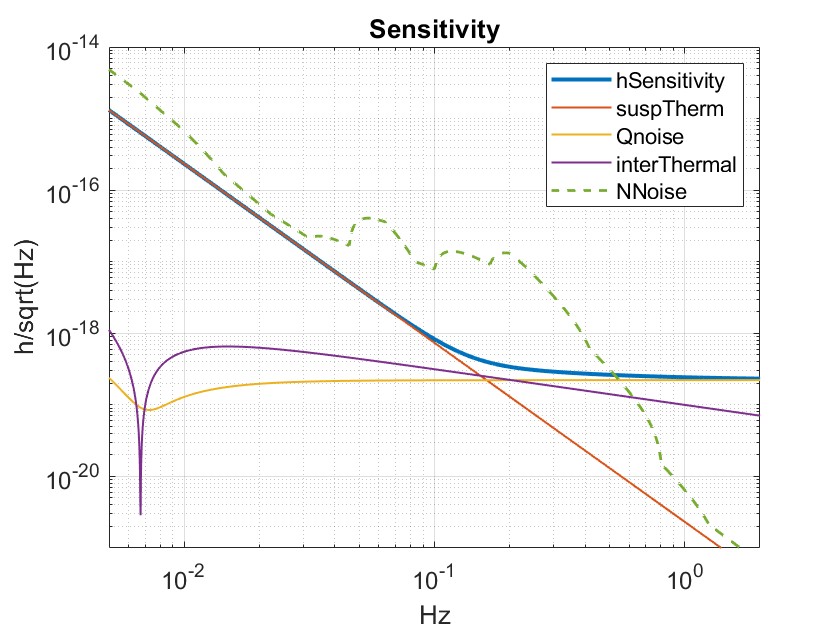}
    \caption{Sensitivity of the detector}
    \label{fig:sensitivity}
\end{figure}
The detector's achievable sensitivity in h, is reported in Fig.~\ref{fig:sensitivity}, where we also estimate the Newtonian Noise, which is expected to be extremely significant and limiting for ground-based detectors at these frequencies. This noise obviously depends heavily on the site, depth, and any active noise reduction methods that may be implemented in future years. Our projection is based on the Peterson New Low Noise Model of seismic noise \cite{peterson}. This is the convolution curve of seismic noise minima recorded at various sites around the world. It may therefore sound optimistic. However, it should be noted that for the Newtonian noise estimation, we do not take into account any reduction, which is currently an extremely active field of research.
The theoretical assessment of Newtonian noise is still being refined and is not yet supported by experimental evidence. For our assessment, we used Eq. 40 in \cite{Vetrano:2013qqa}: 
\begin{equation}
\tilde{h} = 16*\sqrt{2\pi/15} \frac{G \rho_0}{2\pi f C_L} \tilde{\xi};
\end{equation}

where $\rho_0 = 2700 \frac{kg}{m^3}$ is the rock (granite) density and $C_L = 6000 m/s$ is the speed of longitudinal seismic waves. 
It is valid in the approximation of a gravimeter length L  much smaller than the length of the seismic wave, a condition verified in our case across the entire frequency band. 

Being dependent on Newtonian noise, the sensitivity of this instrument will depend both on its location and on the noise reduction techniques that will be developed in the coming years \cite{harms1,harms2,harms3,harms4}.
However, in our opinion, a remarkable quality of the detector is its relative simplicity of construction and sensitivity, comparable to both future ground-based detectors, such as torsion pendulums or those based on atomic interferometry, and space-based detectors, such as LISA, particularly in the frequency range of hundreds mHz.
Compared to these instruments, the solution proposed here can likely be considered complementary, because in the future, having a network of coordinated antennas, even if based on different detection principles, could be an asset.

\begin{acknowledgments}
We wish to acknowledge M. Maggiore for useful discussions.
\end{acknowledgments}

\bibliography{biblioGrav}

\end{document}